\documentclass{aa}
\usepackage{graphics}

\usepackage{natbib}
\usepackage{graphicx}
\usepackage{amssymb}
\begin{document}

\title{XMM-Newton X-ray spectroscopy of classical T Tauri stars}

\author{J. Robrade and J.H.M.M. Schmitt}
\institute{
Hamburger Sternwarte, Universit\" at Hamburg, Gojenbergsweg 112,
D-21029 Hamburg, Germany}

\authorrunning{Robrade \& Schmitt}
\titlerunning{CTTS in X-rays}
\offprints{J. Robrade}
\mail{jrobrade@hs.uni-hamburg.de}
\date{Received 23 September 2005 / Accepted 13 December 2005}

\abstract{

We present results from a comparative study of XMM-Newton observations of four classical T Tauri stars (CTTS), 
namely \object{BP Tau}, \object{CR Cha}, \object{SU Aur} and \object{TW Hya}.
In these objects coronal, i.e. magnetic, activity and as recently shown, 
magnetically funneled accretion are the processes likely to be responsible for the generation of X-ray emission. 
Variable X-ray emission with luminosities in the order of $10^{30}$\,erg/s is observed for all targets. 
We investigate light curves as well as medium and high-resolution X-ray spectra to determine the
plasma properties of the sample CTTS and to study the origin of their X-ray emission
and its variability. 
The emission measure distributions and observed temperatures differ significantly and the targets
are dominated either by plasma at high densities as
produced by accretion shocks or by predominantly hotter plasma of coronal origin.
Likewise the variability of the X-ray luminosity is found to be generated by both mechanisms.
Cool plasma at high densities is found in all stars with detected \ion{O}{vii} triplet emission, 
prevented only for SU~Aur due to strong absorption.
A general trend is present in the abundance pattern, with neon being at solar value or enhanced while 
oxygen, iron and most other metals are depleted, pointing to 
the presence of the inverse FIP effect in active coronae and possibly grain formation in evolved disks.
We find that both accretion shocks and coronal activity contribute to the observed X-ray emission of the targets.
While coronal activity is the dominant source of X-ray activity in the majority of the CTTS,
the fraction for each process differs significantly between the individual objects. 

}
\maketitle

\section{Introduction}
\label{intro}

Young, late-type pre main-sequence stars, known as T~Tauri stars, are copiously found in or near star 
forming regions.  Historically, they are classified according to their H$\alpha$ equivalent width 
as classical T~Tauri stars (CTTS) with EW$>$10\AA\,\,or otherwise as weak line T~Tauri stars (WTTS). 
The principal physical difference between these two classes is that CTTS are thought to be in an earlier evolutionary stage,
i.e. they still possess a disk containing dust and gas and are accreting matter.
WTTS are thought to be more evolved, to have mainly lost their disks and are approaching the 
main-sequence. 
While CTTS are generally younger with ages of a few up to $\sim$\,10\,Myr
and the overall fraction of CTTS in a given star-forming region decreases with age,
both types of T~Tauri stars are commonly found in the same star-forming regions, 
indicating individual evolutionary time scales. Since the H$\alpha$-line is often time-variable,
a classification solely based on its strength is sometimes misleading and
other or additional spectral features have been used to identify and classify CTTS more reliably. 
The existence of a disk containing a significant amount of matter and ongoing accretion onto 
the host star impose major differences between the two types of TTS, leading to different 
spectral properties which can be observed at different wavelengths.
The effects of the warm and probably structured disk are for instance reflected in the different 
infrared designations of CTTS (Class II) and WTTS (Class III) due to their additional near-infrared emission, 
which is an important indicator for CTTS. 
Consequently, the different criteria and underlying data used for the identification of CTTS may introduce
selection effect in the respective samples. 
 
Both types of T~Tauri stars are well known strong and variable X-ray emitters and large numbers of 
these objects were detected with {\it Einstein} \citep{fei81,fei89} and ROSAT \citep{fei93,neu95}.
Their X-ray emission and the observed flaring was usually interpreted
as a scaled-up version of coronal activity in analogy to cool main-sequence stars. 
This picture is still valid and the X-ray emission from WTTS is thought to originate from coronal activity.
However, T Tauri stars are thought to be at least in their early stages fully convective, therefore the
question arises about the underlying processes for the generation of the magnetic field and the influence of accretion on the
stellar structure and corona.
Further on, the evolutionary stages 
of young stars and their high energy emission is of major importance for the process of planetary 
formation via its effects on chemistry, dust and protoplanets.
A comprehensive review of high energy processes and underlying physics in young stellar objects,
summarising the results of the era prior to {\it Chandra} and XMM-Newton, is presented by \cite{fei99}.
 
The largest sample of pre main-sequence stars are found in the Orion Nebula Cloud.
Using {\it Chandra} ACIS observations,
\cite{fei03} analysed a sample of 500 PMS stars with known basic properties. An even deeper exposure, named the
{\it Chandra} Orion Ultradeep Project (COUP) was presented recently, see e.g. \cite{pre05}, confirming and
extending their results.
The X-ray luminosity of CTTS in this sample was found to be correlated with bolometric luminosity.
No strong correlation between $L_{\rm X}$ and rotation as observed for main-sequence stars was found,
contrary, $L_{\rm X}$ correlates with $L_{\rm bol}$ and stellar mass,
indicating the presence of different, e.g. turbulent, dynamos or of supersaturation effects in solar-type dynamos.
The question, whether the presence of an accretion disk influences the stellar X-ray emission is still under debate.
While \cite{fei03} found no intrinsic differences
of X-ray activity related to the presence of a circumstellar disk, the results found by \cite{pre05}
indicate that accretion diminishes  $L_{\rm X}$ on average and weaken the $L_{\rm X}$/$L_{\rm bol}$ correlation.
A similar influence was found by \cite{ste01} in the X-ray properties of young stars 
in the Taurus-Auriga complex observed with ROSAT PSPC.
The COUP sources are rather consistent with the bulk of their X-ray emission being produced in typical coronae.
In general, magnetic activity in CTTS may also involve the circumstellar material, e.g. via star-disk interaction, 
but the topology of their magnetic fields in virtually unknown; \cite{fav05} present evidence
for magnetic star-disk interaction in the analysis of larger flares on several COUP objects. 

Moreover, recent high-resolution spectroscopy of CTTS
with {\it Chandra} and XMM-Newton, indicated accretion shocks at least as an additional mechanism for the
generation of X-ray emission in CTTS.
In this interpretation X-ray emission is generated by a magnetically funneled accretion stream falling
onto small areas of the stellar surface, producing shocked high density plasma \citep{shu94,cal98}.
Since in magnetospheric accretion the matter falls from several stellar radii at nearly free-fall 
velocity, strong shocks and consequently hot plasma emitting in the X-ray regime are formed.
This contribution is expected to be at energies below 0.5\,keV and is distinguishable from cool coronal plasma by its density;
therefore it can be recognised only through high resolution X-ray spectroscopy.
It is thus not contradictory that the COUP sample based on ACIS data with medium resolution spectroscopy 
and poorer sensitivity at low energies finds accretion to play no significant role.

TW~Hya was the first example of a CTTS, where observational evidence for an accretion scenario could be found in X-ray data. 
Analysis using density sensitive lines of He-like triplets (e.g. \ion{O}{vii}, \ion{Ne}{ix})
indicated very high densities, exceeding by far the densities 
previously found in coronal plasma from any other star \citep{twc,twx}.
These low line ratios were also found in 
the spectra of BP~Tau \citep{bptau}, again strongly supporting an accretion scenario.
However, accretion is only capable to produce plasma with temperatures of a few MK and therefore 
contributes nearly exclusively to the low energy X-ray emission.
The observed flaring and additional hard emission attributed to plasma with temperatures of 
several tens of MK in BP~Tau and other CTTS require the presence of an additional 
X-ray generating mechanism such as magnetic activity.
In this work we use the term `coronal activity', but more complex phenomena like star-disk interactions cannot be ruled out.
While TW~Hya like objects where accretion is thought to be the dominant source of the X-ray emission are probably rare, 
the BP Tau observation revealed that accretion as an additional X-ray generating mechanism may be more common.
Another X-ray generating mechanism, shocks in strong Herbig-Haro outflows that may also produce a soft X-ray excess, 
was proposed for jet-driving CTTS (which resemble Class I protostars) like DG Tau \citep{gue05}. 
We note that this model cannot explain the soft excess in the case of TW~Hya or the other sample CTTS.

We utilise XMM-Newton data from four X-ray bright CTTS
to study the origin of their X-ray emission in accretion shocks vs. coronal activity.
We specifically use medium and high resolution X-ray spectra and spectrally resolved X-ray light curves to investigate
the physical properties of our target stars BP~Tau, CR~Cha, SU~Aur and TW~Hya and 
to check the relevance of these mechanisms for the generation of the
X-ray emission and variability in these young stars.
This work is the first comparative X-ray study of CTTS using high and medium resolution 
spectra so far, therefore it is complementary to large sample studies of a specific star forming region such as COUP.
Moreover, the data of CR~Cha and partly of SU~Aur and BP~Tau is analyzed and presented
here for the first time. Results derived from RGS spectra of BP~Tau and the simultaneous obtained UV data were presented in our 
previous letter \citep{bptau}, while the data on TW~Hya first presented by \cite{twx} 
is re-analysed for purposes of comparison in a manner identical to the analysis of all other 
sources.  The plan of our paper is as follows:
In Sect.\,\ref{tar} we describe the individual targets and previous X-ray results,
in Sect.\,\ref{obsana} the observations and the methods used for data analysis and
in Sect.\,\ref{results} we present and discuss the results subdivided into different physical 
topics, followed by a summary and our conclusions in Sect.\,\ref{summ}.

\section{The CTTS sample}
\label{tar}

BP~Tau is a CTTS with spectral type K5--K7 associated with the Taurus-Auriga star forming region at
a distance of 140\,pc as discussed by \cite{wich98}. Optical and infrared measurements clearly 
show the typical characteristics of excess emission produced by accretion \citep{gul98}.
BP~Tau has a moderate inclination angle of roughly 30\degr\,and its
disk was found to be quite compact \citep{muz03}.  X-ray emission from BP~Tau with 
$L_X\sim10^{30}$\,erg/s was already measured with the {\it Einstein Observatory} \citep{wal81},
the RASS luminosity was determined as $0.7\times10^{30}$\,erg/s \citep{neu95}.
\cite{gul97} presented an analysis of simultaneous optical and ROSAT observations of BP~Tau. 
Finding no correlation between optical and X-ray variability they concluded 
that the optical emission is attributed to accretion while the X-ray emission arises from magnetically 
active regions. Initial results of our XMM-Newton observation of BP~Tau with special focus on the
RGS data and density diagnostics with the He-like triplet of \ion{O}{vii}
were presented in our previous letter \citep{bptau}, where evidence for high density plasma and/or 
intense UV radiation is presented, supporting the presence of an accretion funnel in BP~Tau and a rough
estimate of BP~Tau's mass accretion rate as derived from simultaneous UV measurements was given.

\medskip

CR~Cha with spectral type K2 is located in the Chamaeleon I cloud, 
a star forming region at a distance of 140--150\,pc with a measured 
RASS luminosity of $1.5\times10^{30}$\,erg/s \citep{fei93}.
It was classified as CTTS via H$\alpha$, see e.g. \cite{rei96}, and according to \cite{meeus03} it is
dominated by small amorphous silicates, indicating large amounts of unprocessed dust. 
The archival XMM-Newton data of CR~Cha is analysed and
presented here for the first time. 

\medskip

SU~Aur is classified as G2 subgiant and also a member of the Taurus-Auriga star forming region,
therefore we adopt a distance of 140\,pc. For this very luminous and variable object
also a wider range of values of its optical properties is given in literature, see e.g. \cite{dew03}.
It was sometimes classified as WTTS, but a clear IR-excess confirms its status as a CTTS, 
viewed nearly edge on, with a rather high accretion luminosity \citep{muz03}, 
thus causing the strongest absorption signatures in our sample.
SU~Aur is one of the brightest known CTTS, its strong X-ray emission was already detected 
with the {\it Einstein Observatory} \citep{fei81} with an X-ray luminosity of
$3.0\times10^{30}$\,erg/s. A similar value ($3.7\times10^{30}$\,erg/s) was derived from RASS data 
by \cite{neu95}, making it the brightest target in our sample, especially when considering the 
unabsorbed X-ray luminosity.  Some results of the XMM-Newton observation are presented in an analysis 
of star forming regions by \cite{pal04}, who showed that SU~Aur
is both extremely hot and variable.

\medskip

TW~Hya is classified as a K7--K8 star and is the nearest object in our sample. 
At a distance of only 56\,pc it is located
in the TW Hydra association, a nearby young but diffuse stellar associations, see e.g. \cite{zuck01}.
TW~Hya is viewed nearly pole on and due to its proximity and lack of obscuring clouds
it can well studied at all wavelengths. With an estimated age of nearly 10\,Myr it is
also one of the oldest known stars, still accreting matter.
\cite{kas99} analysed ASCA and ROSAT data of TW~Hya and found 
that a model with plasma temperatures of $\sim$\,1.7 and $\sim$\,9.7\,MK and an X-ray luminosity 
of $\sim\,2\times10^{30}$\,erg/s describes both observations well. 
TW~Hya was also observed with the {\it Chandra} HETGS detector \citep{twc}, 
providing high resolution spectra, which were modelled
with an iron depleted and neon enhanced plasma and a emission measure distribution showing a sharp peak around 3\,MK. 
The analysis of density sensitive lines indicated plasma at
high densities (log n$_e$=13) and the inferred plasma properties were found to be consistent with the X-ray
emission to be generated by funneled mass accretion from the circumstellar disk.
Several moderate 'flares' were detected on TW~Hya during the ASCA and {\it Chandra} observation;
however, the untypical shape of the flare light curve
and the seemingly constant spectral properties led \cite{twc} to conclude that most or all of the
X-ray emission of TW~Hya originates from accretion.
Results of the XMM-Newton observation of TW~Hya obtained with a somewhat different analysis were 
presented by \cite{twx}, who found that the XMM-Newton 
data can be well explained by X-ray emission from a metal depleted accretion shock with 
an X-ray luminosity of $\sim\,1.5\times10^{30}$\,erg/s, 
consistent with the {\it Chandra} results. 
Different conclusions were derived at on the interdependent properties
mass accretion rate and surface area of the shock region, 
which is assumed to fill either below one percent or up to a few percent of the stellar surface.
Here the TW~Hya data is independently re-analysed to ensure consistency and allow a comparison of our results.

\begin{table}[!ht]
\caption{\label{ctts}Basic properties for our sample CTTS. Spectral types taken from Simbad database,
distances adopted from \cite{wich98}, ages from \cite{gul98}:\,BP\,Tau, \cite{nat00}:\,CR\,Cha, \cite{dew03}:\,SU\,Aur, 
\cite{mak01}:\,TW\,Hya, $L_{x}$ are ROSAT values ($<$\,2.4\,keV).}
{\scriptsize
\begin{tabular}{lcccc}\hline\hline
Target & Spec.T. &  Dist.(pc)  & Age (yr)& $L_{x}(10^{30}erg\,s^{-1})$\\\hline
BP Tau & K5V & 140 & $6\times10^5$ & 0.7\\
CR Cha & K2 & 145 &  $1\times10^6$ & 1.5 \\
SU Aur & G2III & 140 &  $4\times10^6$ & 3.7 \\
TW Hya & K8V & 56& $8\times10^6$ & 2.0 \\\hline
\end{tabular}
}
\end{table}

\section{Observation and data analysis}
\label{obsana}

All sample CTTS were observed with XMM-Newton using somewhat different detector setups 
with exposure times in the range of 30\,--\,130\,ks. 
Data were taken with all X-ray detectors, which were operated simultaneously onboard XMM-Newton, 
respectively the EPIC (European Photon Imaging Camera), consisting of the MOS and PN detectors
and the RGS (Reflection Grating Spectrometer). Note that for SU~Aur no PN data is available and
different filters were used, the thick filter for the BP~Tau and SU~Aur
and the medium filter for the CR~Cha and TW~Hya observations.
Further on, the signal to noise ratio differs for the various targets and instruments.
For CR~Cha only a fraction of the original data from the additionally split observation
could be used for analysis because of very high background contamination; 
the RGS data from two exposures was merged with the tool 'rgscombine' provided with SAS\,6.5, but 
the RGS spectrum of CR~Cha is still underexposed. 
The EPIC data is quality is sufficient for all targets.
In addition, strong absorption affects the data quality especially at lower energies. 
The details of our observations and used data are described in Table~\ref{obs},
a detailed description of the XMM-Newton instruments can be found in \cite{xmm}.

\begin{table}[!ht]
\caption{\label{obs}Observation log of our sample CTTS, duration of prime instrument/RGS(filtered).}
{\scriptsize
\begin{tabular}{lccc}\hline\hline
Target & Obs.Mode &  Obs. Time  & Dur. (ks)\\\hline
BP Tau & FF/thick & 2004-08-15T06:14--16T18:51 & 132/124 \\
CR Cha & FF,LW/med. & 2001-02-24T05:04--15:49 & 39/72 \\
SU Aur & FF/thick & 2001-09-21T01:27--22T14:18 & 130/126 \\
TW Hya & FF/med. & 2001-07-09T05:51--14:01 & 30/28 \\\hline
\end{tabular}
}
\end{table}

Data analysis was performed with the XMM-Newton Science Analysis System (SAS) software, version 6.0. Images, light 
curves and spectra were produced with standard SAS tools and standard selection criteria were applied for filtering 
the data, see \cite{sas}.
X-ray spectral analysis was carried out with XSPEC V11.3 \citep{xspec}, while
for line fitting purposes we used the CORA program \citep{cora}, assuming Lorentzian line shapes.
Individual line fits are used to investigate the density sensitive forbidden and intercombination 
lines of He-like triplets. Line counts are derived keeping the line spacing fixed within a triplet 
and using an overall line width for all lines.
Spectral analysis of EPIC data is performed in the energy band between 0.3\,--\,10.0\,keV, while the 
RGS first order spectra in the full energy range, 
i.e. 0.35\,--\,2.5\,keV (5\,--\,35\,\AA), are used whenever data quality permits.
While the RGS obviously has the highest spectral resolution,
the EPIC detectors are more sensitive and are able to measure higher energy X-rays; 
the MOS detectors provide a slightly better spectral and spatial resolution, the PN detector is more sensitive.  
We emphasize that 
the data were analysed simultaneously but not co-added, thus providing sufficient signal
for all observations and ensuring a consistent analysis for all targets. 
All periods affected by proton flares were removed from spectral analysis
and for TW~Hya the PN data was cleaned for some moderate pile-up.
The background was taken from source free regions on the detectors. 

For the analysis of the X-ray spectra we use multi-temperature models with variable but tied abundances,
i.e. the abundance pattern was assumed to be the same in all temperature components.
Such models assume the emission spectrum of a collisionally-ionized optically-thin gas 
as calculated with the APEC code, see e.g. \cite{apec}. 
Since the RGS is not very sensitive above 2.0\,keV, 
we use EPIC data to determine the properties of plasma with temperatures above 20\,MK. 
In order to account for calibration uncertainties of the different
detectors, see e.g. \cite{kir04}, the normalization between instruments was taken as a free parameter for each type of
instrument, i.e. for MOS, PN, RGS. 
Our fit procedure is based on $\chi^2$ minimization, therefore spectra 
are always rebinned to satisfy the statistical demand of a minimum value of 15 counts per spectral bin.
All errors are statistical errors given by their 90\% confidence range and were calculated separately
for abundances and temperatures by allowing variations of normalizations and respective model parameters.  
Note that additional uncertainties arise from uncertainties in the atomic data and 
instrumental calibration which are not explicitly accounted for.

The applied models use three temperature components,
models with additional temperature components were checked, but did not improve the fit results significantly.
Since the APEC models do not account for high density plasma,
resolved density sensitive lines are excluded from the global fits for TW~Hya.
Abundances are calculated relative to solar photospheric values as given by 
\cite{and89}. For iron and oxygen we use the updated values of \cite{grev98}.
Application of the new but controversial
solar abundances published by \cite{asp05} would further increase the neon abundance in our sample stars
compared to other metals, but not change our conclusions significantly.
We note that relative results between the CTTS are not affected by the underlying solar abundance pattern.
For elements with overall low abundances and no significant features in the X-ray spectra, 
i.e. Al, Ca, Ni, the abundances were tied to the iron abundance.
When data quality permits, we determine the abundances of individual elements,
global abundances are used if no features are present in the spectra.
This is for instance the case for targets with strong absorption, where e.g. the stronger carbon lines at the
low energy end of the detectors are completely absorbed.
We simultaneously modelled the temperatures and 
emission measures (EM=$\int n_{e}n_{H}dV$) of the components and checked the derived results for consistency.
X-ray luminosities were then calculated from the resulting best fit models.

Absorption in the circumstellar environment and possibly also in the interstellar medium is significant for CTTS
and is applied in our modelling. 
Since absorption is supposed to be slightly variable and, in addition,
optical and IR measurements often give a wider range of values, it is kept as a free parameter in our analysis.
In general, the derived fit results are quite stable, but note that some of the
fit parameters are mutually dependent, thus affecting especially absolute values.
Interdependence mainly affects the low energy region of the spectra, where the strength of absorption, 
the emission measure and abundances 
of elements with emission lines in the respective temperature range strongly depend on each other.
While some effects can be prevented by the methods described above,
models with different absolute values of the mentioned parameters but only marginal differences in its statistical
quality may be fitted to the data, however ratios and relative changes of these properties are again very robust.

\section{Results}
\label{results}

\subsection{X-ray light curves}
\label{lcana}

\begin{figure}[!ht]
\vspace*{-13mm}
\includegraphics[width=90mm]{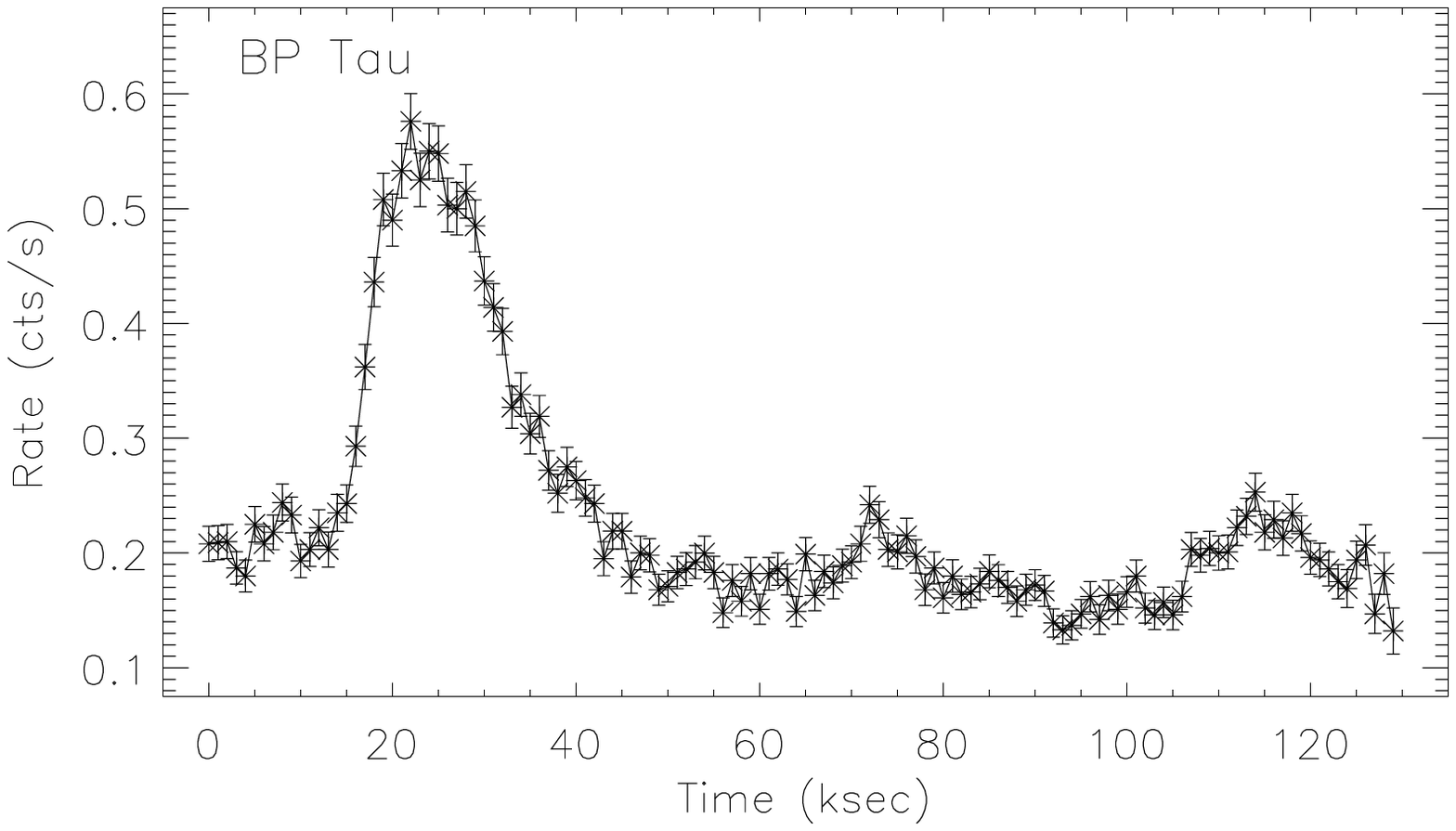}

\vspace*{-15mm}
\includegraphics[width=90mm]{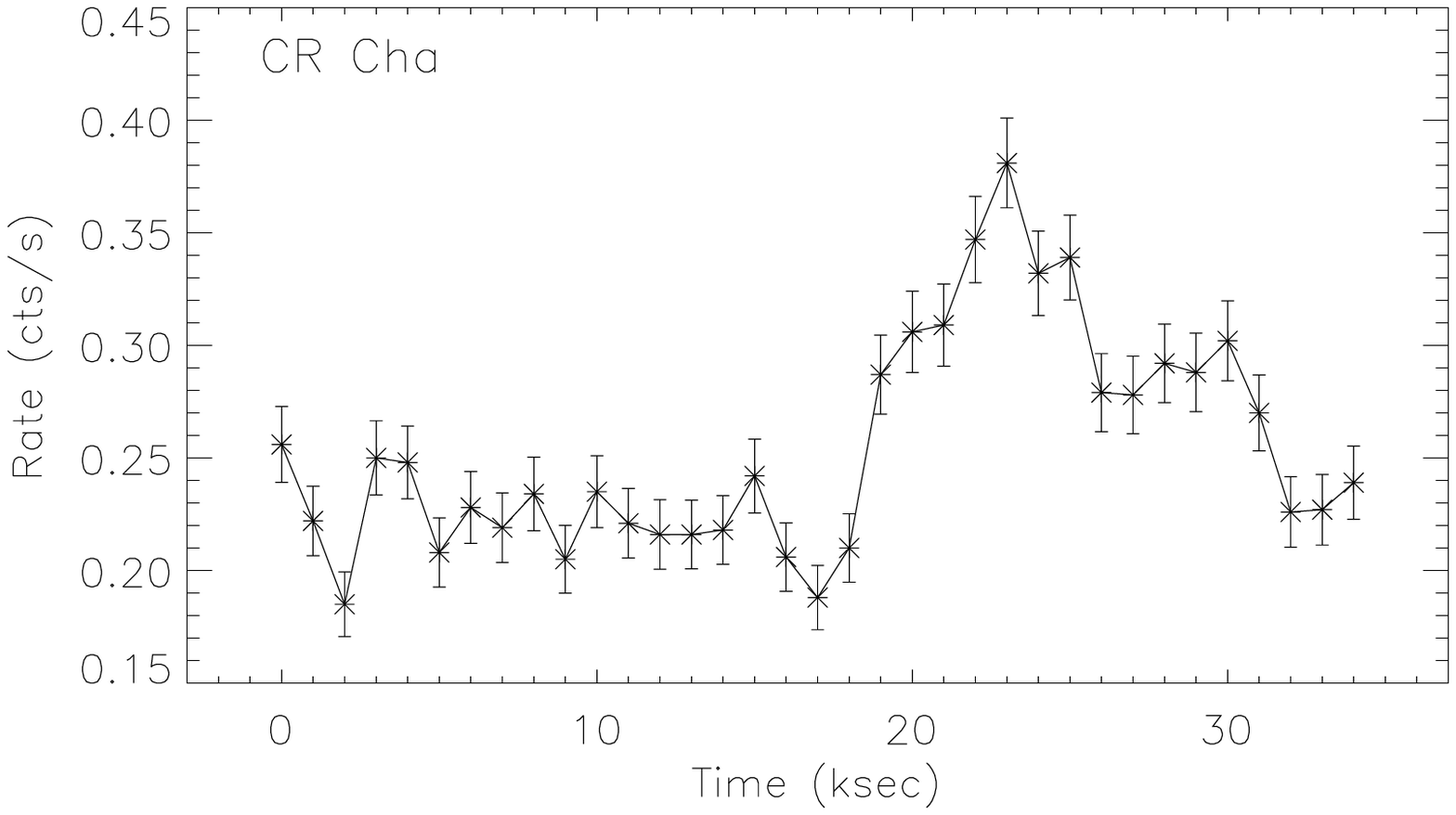}

\vspace*{-15mm}
\includegraphics[width=90mm]{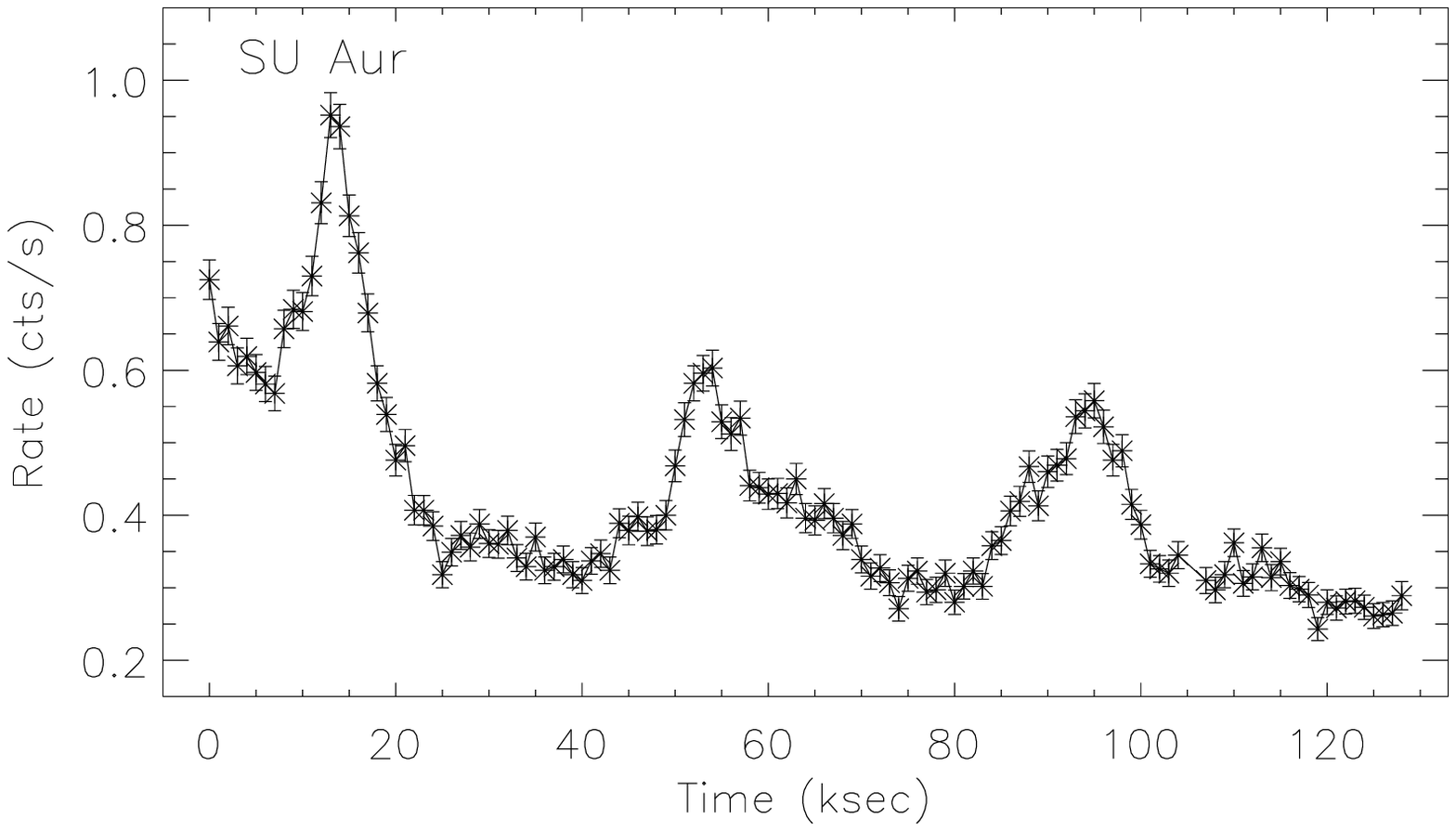}

\vspace*{-15mm}
\includegraphics[width=90mm]{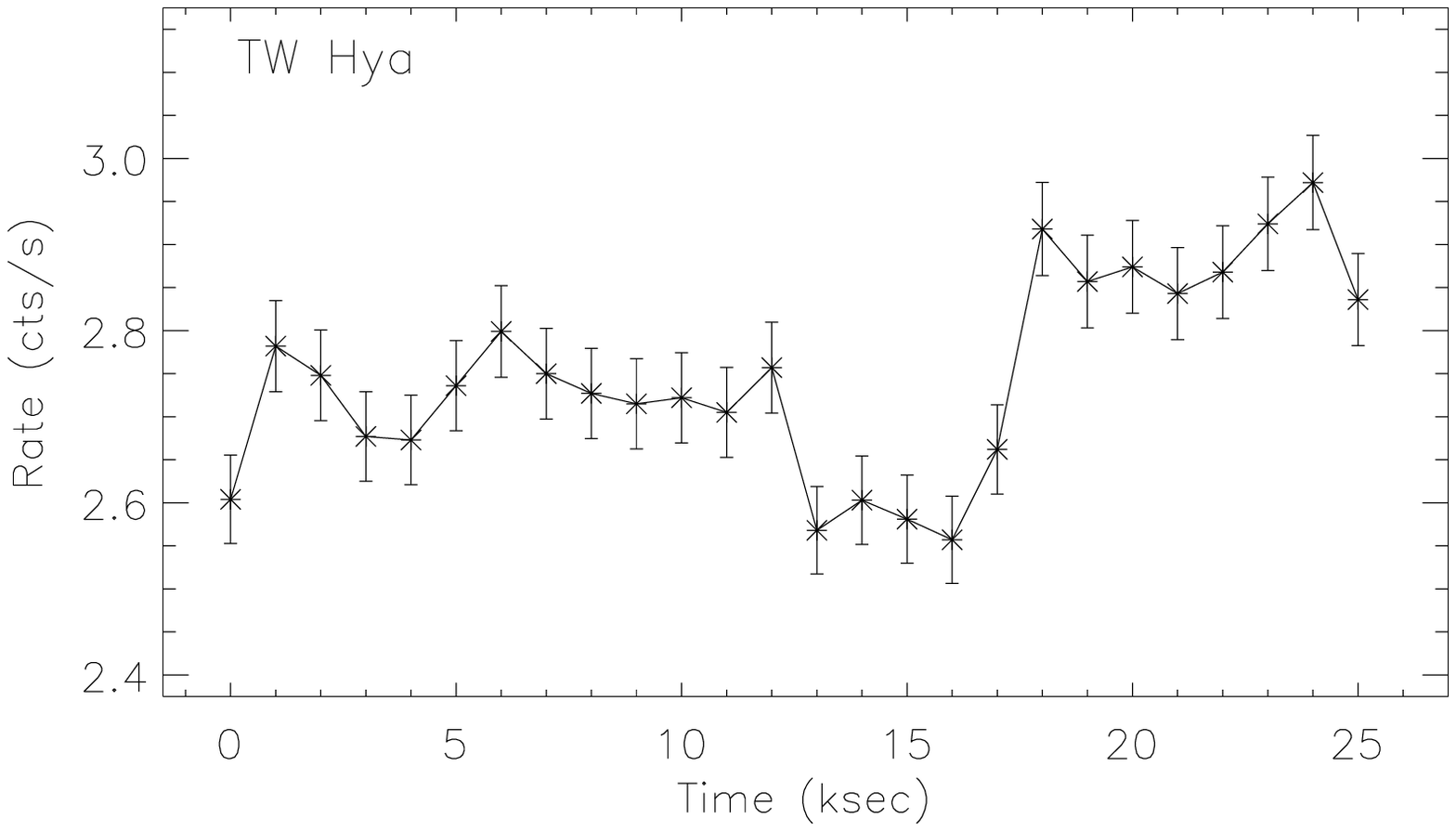}
\caption{\label{lcs}Light curves and corresponding hardness ratio of the sample CTTS, 
PN (SU~Aur\,--\,MOS) data with 1\,ksec binning.}
\end{figure}

To study time variability and its origin we first investigated the X-ray light curves of our sample CTTS.
In Fig.\ref{lcs} we show background subtracted light curves with a temporal binning of 1000\,s,
extracted from a 50\,\arcsec\,radius circular region around each source and
cleaned for obvious data dropouts.

A flare with a rise in count rate of factor 2.5\,--\,3 occurred during the BP~Tau observation
and a smaller flare (factor $\sim$\,1.5) during the CR~Cha observation. In the SU~Aur data several flares 
were detected, the largest one showing an increase in X-ray brightness by a factor of 3.
TW~Hya does not exhibit a flare during the XMM-Newton observation, any variability remains in the range of only 10\%.
However, light curve variations with factors around two are also known for TW~Hya \citep{kas99,twc}, who
noted three 'flares' during 140\,ks (ASCA,\,94\,ks\,\,+{\it Chandra},\,48\,ks) observation time
but no spectral changes were detected during the periods of increased X-ray brightness.
The average time scale for flares on CTTS appears to be around one
moderate flare (factor $\gtrsim$\,2) occurring daily on our sample CTTS.
In addition to obvious flare events,
variability is present also on smaller time scales and lower amplitude variations throughout the observations.

\subsubsection{X-ray hardness}

\begin{figure}[!ht]
\includegraphics[width=90mm,height=80mm]{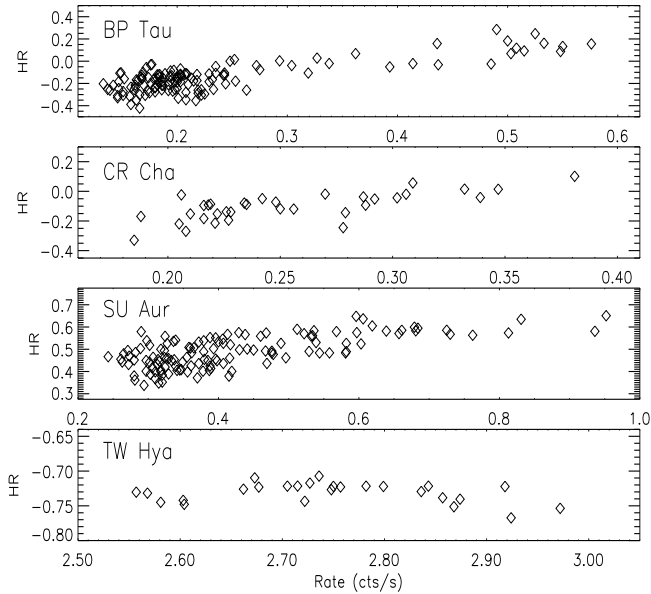}
\caption{\label{hr}Hardness ratio vs. count rate of the sample CTTS, PN (SU~Aur\,--\,MOS) data with 1\,ksec binning.}
\end{figure}

The light curves in connection with spectral hardness can be used to identify the origin of the variability.
While in typical stellar coronal flares the emission measure of predominantly hot plasma
and hence the hardness of the spectra is increased,
in a pure accretion spectrum no spectral changes should accompany the brightening, 
since the plasma temperature only depends on the infall velocity and not the accretion rate. 
If, in addition, a coronal contribution with a temperature higher than that produced by accretion is present, 
a slight spectral softening should be observed.
We calculate a hardness ratio
for each light curve time bin and in Fig.\,\ref{hr} we show the hardness ratio vs. count rate for our sample CTTS.
The hardness ratio is here defined as HR=H-S/H+S with
the soft band covering the energy range 0.2\,--\,1.0\,keV and the hard band 1.0\,--\,10.0\,keV.
Errors are small compared to the observed shifts in hardness ratio.
A clear correlation of X-ray brightness with spectral hardness is present for BP~Tau, CR~Cha and SU~Aur, 
a behaviour typical of stellar flares, suggesting a coronal origin of the variability.
No strong correlation is found for variations on TW~Hya, which are, however, quite small.
An anti-correlation appears to be present for the larger variations, 
as expected for brightness changes due to an increase in accretion rate.
This is reflected in the slope of the linear regression curves. 
It is positive for BP~Tau, CR~Cha and SU~Aur, but slightly negative for TW~Hya.

To quantify the spectral changes that the CTTS are undergoing during phases of increasing emission in more detail,
we use their light curves to define 'high' states for BP~Tau (15--37\,ks), CR~Cha (18.5--31.5\,ks), 
SU~Aur (0--24, 47--69, 83--101\,ks) and TW~Hya ($>$ 19\,ks), while the remaining time intervals are considered as 
quasi-quiescence. An individual spectral analysis of both states of activity is presented in Sect.\,\ref{specch}.

\subsection{Spectral analysis}
\label{specana}

The spectral analysis and its results are presented as follows.
First we use the medium resolution EPIC spectra in combination with the high resolution RGS data
to obtain general spectral properties like temperatures, emission measures 
and elemental abundances of the X-ray emitting plasma and constrain the strength of
absorption present for our targets.
From the RGS spectra we investigate the resolved He-like triplet of \ion{O}{vii}
to determine the densities of the emitting plasma and/or strength of the surrounding UV field.
Subsequently we investigate spectral changes between the low and the high state of our sample CTTS.

\subsubsection{The global spectra}
\label{specg}

\begin{figure}[ht]
\includegraphics[width=50mm,angle=-90]{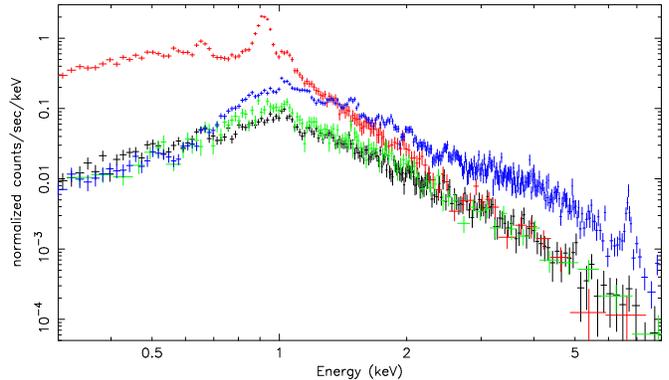}
\caption{\label{mos1}Spectra of our sample CTTS for the total observation as observed with the MOS1 detector: 
BP~Tau (black), CR~Cha (green), SU~Aur (blue) and TW~Hya (red); bottom to top of the 1.0\,keV peak.
{\it Colour figure in electronic version.}}
\end{figure}

The spectra of all our sample CTTS as observed with MOS1 are shown in Fig.\,\ref{mos1}, 
which demonstrates major differences between individual stars. 
While the spectra of BP~Tau and CR~Cha are comparable, the spectrum of SU~Aur indicates 
the presence of large amounts of extremely
hot plasma noticeable, e.g. in the very strong \ion{Fe}{xxv} line complex at 6.7\,keV.
On the other hand, the X-ray spectrum of TW~Hya is much softer, suggesting a 
more dominant accretion component. The observed spectra are also
subject to absorption, affecting primarily the energy range below 1.0\,keV. In these spectra the low energy slope mainly
reflects the strength of the absorption, while the high energy slope traces temperature and amount of
hot coronal and flaring plasma. Inspection of the two slopes indicates that the
absorption is weakest for TW~Hya, moderate for BP~Tau and strongest for CR~Cha and SU~Aur, while
the coronal component is strongest and hottest for SU~Aur, followed by BP~Tau and CR~Cha 
and much weaker and cooler for TW~Hya.

To quantify the physical properties of the X-ray emitting plasma we fitted the spectra of the
different instruments with multi-temperature spectral models as described in Sect.\ref{obsana}.
The results of the fitting procedure are presented in Table\,\ref{specres}. 
Note again the interdependence of several fitting parameter,
especially absorption, cool emission measure and corresponding abundances.
However, general trends on abundances and their ratios as well as the overall shape 
of the emission measure distribution and their changes,
which indicate the production mechanism of the X-rays, are very stable.

While spectral modelling must consider absorption, the physically most interesting quantity is 
the emitted, i.e. dereddened, X-ray luminosity 
of our targets, which is calculated in the 0.2--10.\,keV band using MOS data (see \ref{specres}). 
We find that despite their very different spectral properties,
all targets have an X-ray luminosity of the same order of magnitude.
The hottest and most massive star SU~Aur is also X-ray brightest target, followed by CR~Cha, BP~Tau and 
TW~Hya with comparable X-ray luminosities.
CTTS as a class are known strong X-ray emitter, but our sample stars are about a magnitude brighter than the 
X-ray luminosity of an average CTTS, as determined from RASS data 
with a mean log$L_{X}$ of 29.1 (Chamaeleon) and 29.4 (Taurus) in \cite{neu95}, \cite{fei93}; 
this is of course no surprise since these CTTS were selected for grating observations.

\begin{table}[!ht]
\setlength\tabcolsep{5pt}
\caption{\label{specres}Spectral results, units are $N_{H}$ in $10^{21}$cm$^{-2}$, kT in\,keV EM in $10^{52}$cm$^{-3}$
and $L_{\rm X}$ in $10^{30}$\,erg\,s$^{-1}$.}
\begin{center}
{\scriptsize
\begin{tabular}{lcccc}\hline\hline
Par. & BP Tau & CR Cha & SU Aur & TW Hya\\\hline
$N_{H}$ & 1.5$^{+ 0.1}_{- 0.1}$ & 2.7$^{+ 0.4}_{- 0.3}$&  3.1$^{+ 0.1}_{- 0.1}$ & 0.35$^{+ 0.04}_{- 0.05}$\\
Fe &0.28$^{+ 0.04}_{- 0.04}$ & 0.45$^{+ 0.08}_{- 0.07}$ &  0.79$^{+ 0.05}_{- 0.05}$  &0.26$^{+ 0.02}_{- 0.02}$\\
Si &0.14$^{+ 0.08}_{- 0.08}$ & 0.23$^{+ 0.10}_{- 0.09}$ &  0.60$^{+ 0.08}_{- 0.08}$  &0.13$^{+ 0.05}_{- 0.06}$ \\
O &0.62$^{+ 0.08}_{- 0.06}$ & 0.53$^{+ 0.18}_{- 0.14}$ &  0.39$^{+ 0.12}_{- 0.11}$  &0.23$^{+ 0.02}_{- 0.02}$\\
Ne &1.47$^{+ 0.16}_{- 0.14}$ & 1.02$^{+ 0.19}_{- 0.17}$ &  1.04$^{+ 0.15}_{- 0.14}$  &1.81$^{+ 0.05}_{- 0.05}$\\
kT1 & 0.20$^{+ 0.01}_{- 0.01}$ & 0.17$^{+ 0.03}_{- 0.03}$ & 0.68$^{+ 0.03}_{- 0.02}$   &0.25$^{+ 0.01}_{- 0.01}$\\
kT2 & 0.63$^{+ 0.03}_{- 0.03}$ & 0.66$^{+ 0.01}_{- 0.02}$ & 1.61$^{+ 0.07}_{- 0.10}$   &0.69$^{+ 0.06}_{- 0.02}$\\
kT3 & 2.17$^{+ 0.09}_{- 0.09}$ & 1.92$^{+ 0.06}_{- 0.10}$ & 4.82$^{+ 1.02}_{- 0.84}$   &1.34$^{+ 0.03}_{- 0.05}$\\
EM1 & 3.48$^{+ 0.89}_{- 0.69}$& 4.96$^{+ 4.05}_{- 2.49}$ &  11.66$^{+ 0.87}_{- 0.86}$  &19.16$^{+ 0.74}_{- 0.82}$\\
EM2 & 5.28$^{+ 0.64}_{- 0.56}$& 10.67$^{+ 1.60}_{- 1.53}$&  25.27$^{+ 2.87}_{- 3.96}$  &1.23$^{+ 0.54}_{- 0.84}$\\
EM3 & 10.30$^{+ 0.43}_{- 0.29}$ & 9.27$^{+ 0.73}_{- 0.54}$& 15.79$^{+ 4.02}_{- 2.59}$ &3.12$^{+ 0.52}_{- 0.51}$\\\hline
$\chi^2${\tiny(d.o.f.)} & 0.97 (1281) & 1.20 (700) & 1.20 (1027)  &1.57 (878)\\\hline\hline
$L_{\rm X}$ obs. & 1.3 & 1.4 & 4.5 & 1.5\\
$L_{\rm X}$ emit.& 2.3 & 3.1 & 8.1 & 2.0\\\hline
\end{tabular}
}
\end{center}
\end{table}

Neglecting interstellar contribution, the strength of absorption depends
on the amount of circumstellar material in the disk and the inclination angle of the target. 
The results of our spectral modelling confirm that absorption is lowest for TW~Hya. 
Our value for the interstellar column density $N_{H}$ is slightly higher than the one used by \cite{twx}, 
who fixed $N_{H}$ at $2\times10^{20}\,cm^{-2}$.
No values are given for the {\it Chandra} data in \cite{twc}, but the authors refer to previous work \citep{kas99} where
values determined from ROSAT PSPC ($0.5\times10^{21}\,cm^{-2}$) and higher ones from
ASCA SIS ($2.9\times10^{21}\,cm^{-2}$) with a consequently larger emission measure.
These differences were attributed to temporal changes in the circumstellar environment,
but the poorer spectral resolution and sensitivity of the used instruments, the strong interdependence 
between $N_{H}$ and EM, that is also present in models from the individual instruments operated simultaneously 
onboard XMM-Newton, suggests that limited spectral resolution of the instruments 
combined with uncertainties in the code used for modelling the spectra is a possible explanation for those discrepancies.
Recently \cite{twuv} analysed the Ly\,$\alpha$ profile of TW~Hya as measured by HST/STIS and put a stringent 
upper limit on the hydrogen column density with $N_{H}\le5\times10^{19}\,cm^{-2}$. 
We therefore adopted this value and additionally fitted the spectra
with the reduced absorption value, which results in a comparable model, 
but with accordingly lower emission measure at cool temperatures.

We find intermediate values of the interstellar column density $N_{H}$ for BP~Tau, 
which are in good agreement with optical measurements 
using the standard conversion between $N_{H}$ and $A_{v}$ viz. $N_{H}$=$2\times10^{21}A_{v}\,cm^{-2}$;
values taken from literature are in the range $1.0-2.0\times10^{21}\,cm^{-2}$ \citep{gul98,har95}.
Even stronger absorption is found for CR~Cha, also in good agreement with optical measurements,
with a literature value of $2.7\times10^{21}\,cm^{-2}$ \citep{gau92}.
The strongest absorption is found for SU~Aur, here the large inclination angle
results in a strong and probably also variable absorption. Optical and NIR measurements give a range of values 
ranging from no absorption up to $3\times10^{21}\,cm^{-2}$ \citep{dew03}. While commonly a lower $N_{H}$ value
is adopted, our result is comparable with the larger values and a much weaker absorption significantly worsens the
quality of the fit. 

The emission measure distributions (EMD) of our sample CTTS differ significantly among the sample stars.
The hottest EMD is clearly present in SU~Aur, with
significant amounts of plasma at temperatures around 50\,MK 
detected. Due to the very strong absorption no definite 
conclusions about a possibly existing cool temperature part of the EMD can be drawn.
Since SU~Aur is of average age in our sample and the major difference to the other stars is its spectral type,
an explanation for its outstanding properties might be a dependence of the evolutionary time scale
on spectral type, i.e stellar mass.
The EMDs of BP~Tau and CR~Cha are similar, with the temperatures of BP~Tau being 
slightly higher and a larger fraction of plasma residing
in the hottest component; plasma at temperatures around 20\,--\,25\,MK must be present in those stars.
The picture is very different for TW~Hya, where the cool component with temperatures around 3\,MK clearly dominates
the EMD. Plasma at medium temperatures with 5\--\,10\,MK appears to be nearly absent, while in the hot component 
plasma temperatures around 15\,MK are reached; however, this component is 
cooler and significantly weaker than in the other objects. 
We emphasise that this finding is independent of the used value of $N_{H}$.

As far as abundances are concerned, one property is common to all CTTS analysed:  
Most abundances are subsolar, sometimes at considerable level, while neon is commonly found
at solar abundance or even significantly enhanced. 
The noble gas neon is more abundant compared to iron, oxygen and silicon for all targets, and
in the most extreme case TW~Hya these ratios are of the order of 10 (see Table.\,\ref{specres}).
Iron and silicon are less abundant than oxygen in BP~Tau and CR~Cha,
while all three elements are strongly depleted in TW~Hya. The very active star SU~Aur appears to be an exception,
with only oxygen apparently being more strongly depleted, 
but here very strong absorption affects the results based on cool lines.
These metal anomalies were interpreted for TW~Hya by \cite{twx}
through a depletion of grain forming element in the accreted material, consistent with the low observed IR excess
and also plausible when considering the age of TW~Hya. \cite{dra05} argue, that for TW~Hya coagulation of grains 
into larger bodies finally withdraws these elements from the accretion process, while for the younger BP~Tau 
dust and grains at corotation radius ($R_{co}=(\frac {G\, M}{\omega ^2})^{1/3}$)
will sublimate and rereleased into the accretion process. The coagulation of significant amounts of 
material into centimeter sized particles in the disk around
TW~Hya is supported by the radiospectrum at centimeter wavelength as observed with the VLA \citep{wil05}.

However, stars without accretion but with an active corona 
also show a distinct abundance pattern, commonly known as IFIP effect, 
i.e. an enhancement of elements with a high first ionisation potential.
While iron and silicon are low FIP elements, oxygen is an intermediate and neon a high FIP element.
The IFIP-effect is observed for BP~Tau and CR~Cha, thus pointing towards the typical abundance pattern for coronal plasma.
Moreover, \cite{twa5} analyse the spectra of TWA5, a young multiple system dominated by a M~dwarf and also located
in the TW~Hya association. These object is classified as CTTS (via H$\alpha$), but is probably a WTTS
since there is no evidence for a disk in the IR, but it also shows a high neon to iron ratio similar to TW~Hya
with spectral properties very much reminiscent of active stars. While environmental conditions may be invoked to explain
the abundance anomalies,
\cite{mdwarfs} showed that a high neon to iron ratio is common for active M~dwarfs, with most 
extreme ratios found in very young and therefore more active stars.
We further point out that our CTTS belong to different star formation regions
and their X-ray emission is predominantly generated by different processes,
yet no significant difference in the abundance pattern was found. 
Therefore it appears reasonable to attribute the observed abundance anomalies in the sample CTTS
to a combined action of metal depletion of the accreted matter via grain forming and the coronal IFIP effect.
Grain forming is only important for the accretion plasma of older and more evolved objects, while the IFIP effect
is present in plasma generated in active coronae.

\subsubsection{The \ion{O}{viii} and \ion{O}{vii} lines}
\label{specox}

\begin{figure}[t]
\includegraphics[height=85mm,width=35mm,angle=-90]{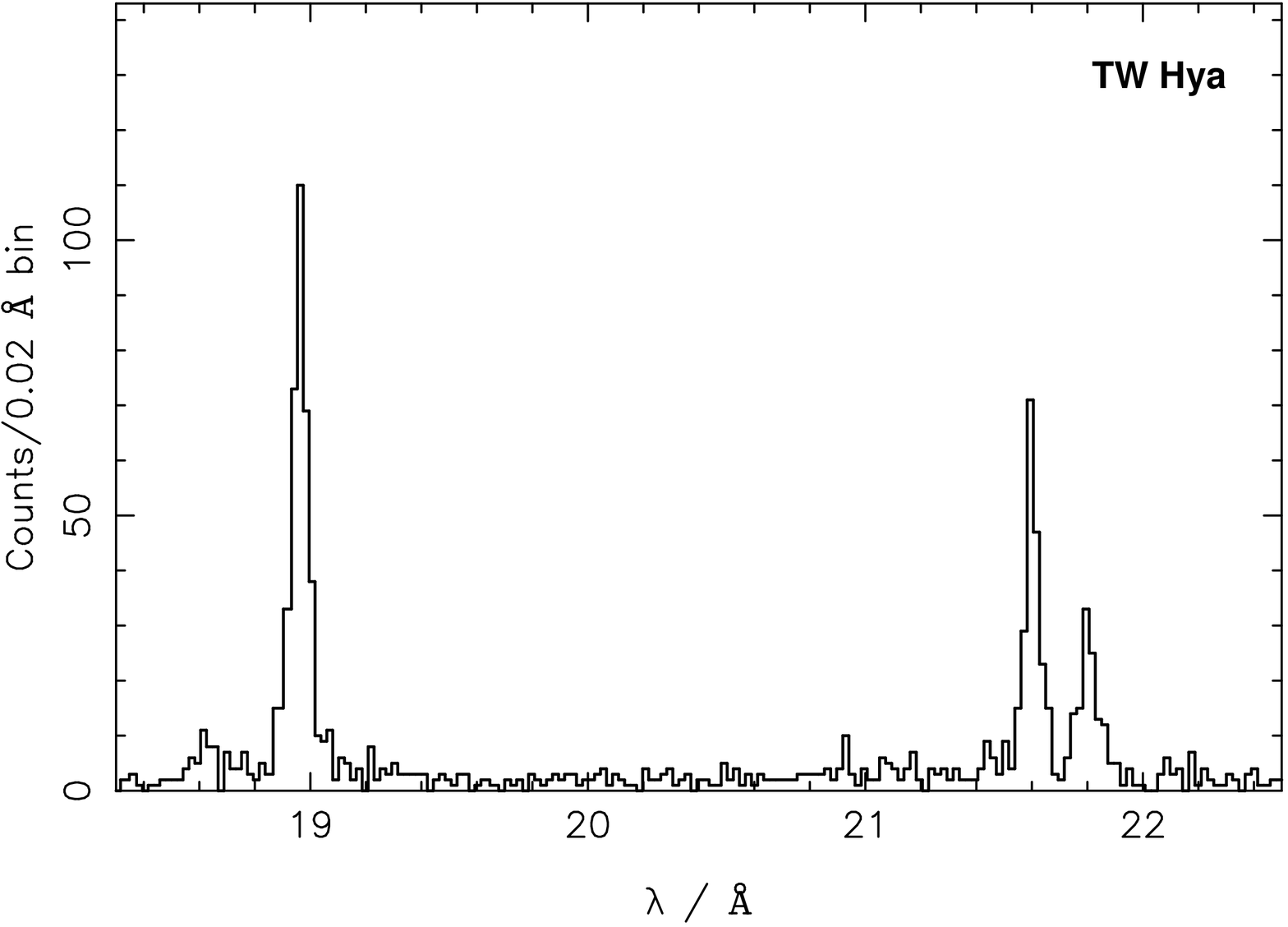}
\includegraphics[height=85mm,width=35mm,angle=-90]{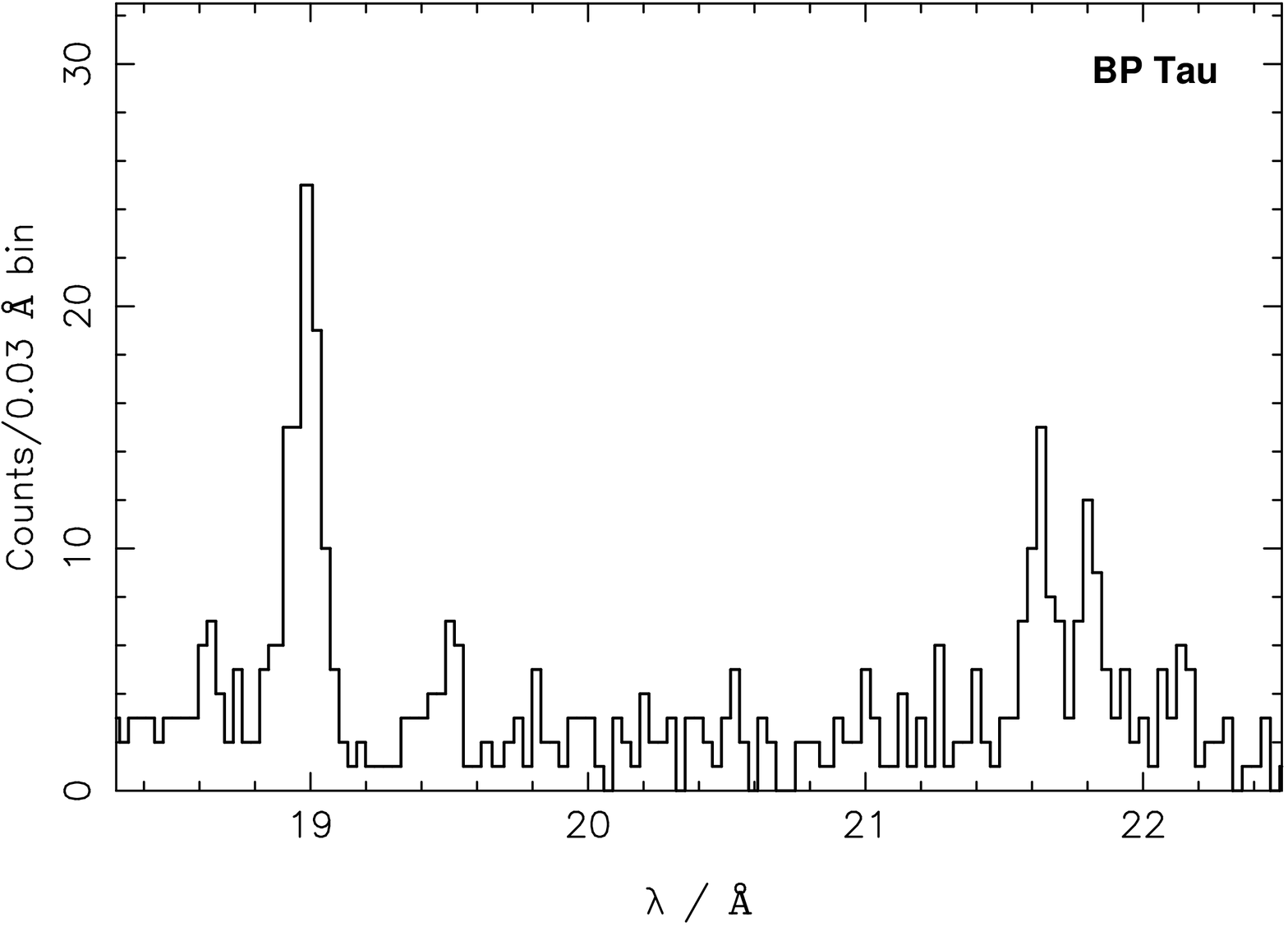}
\includegraphics[height=85mm,width=35mm,angle=-90]{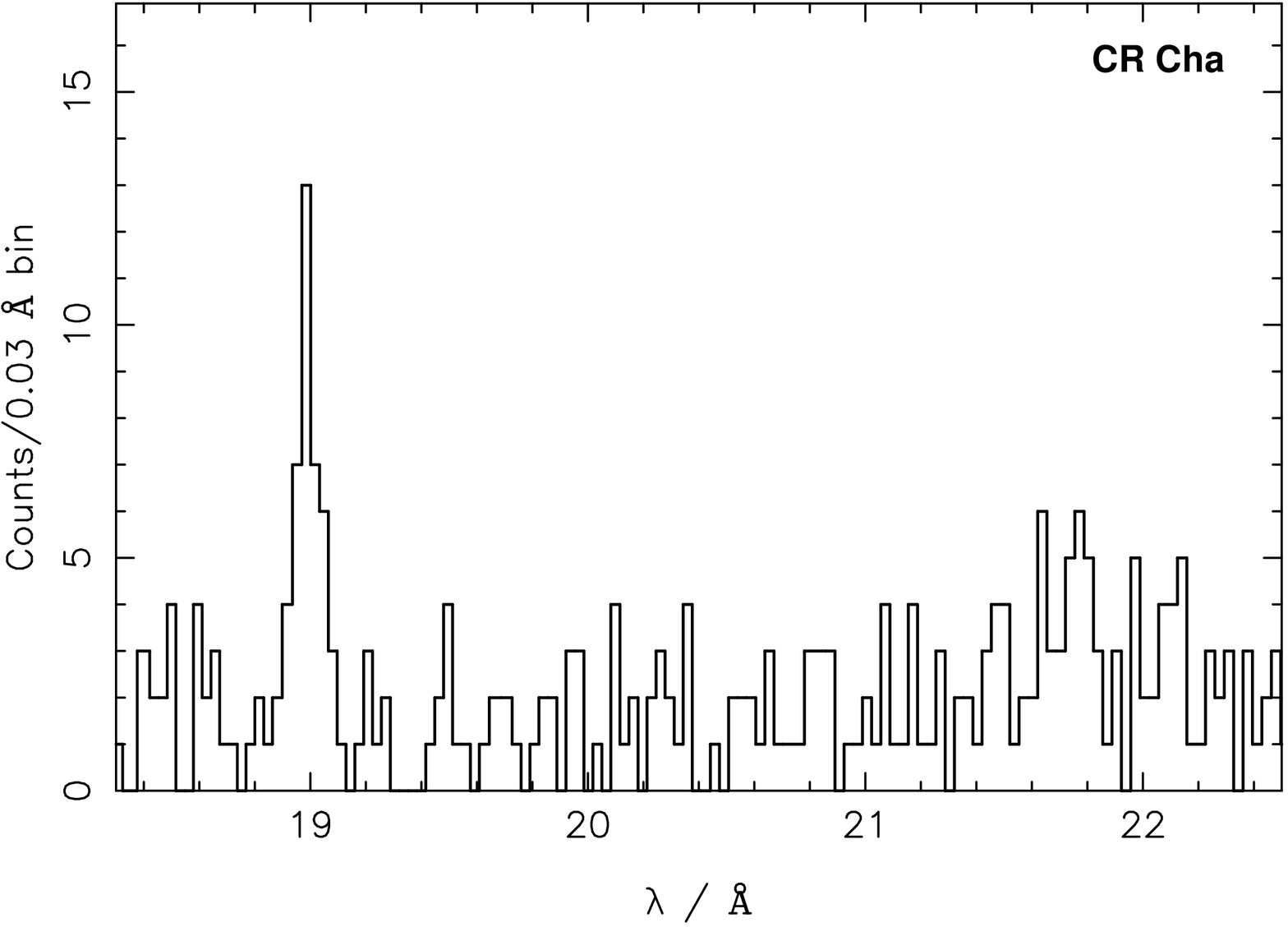}
\includegraphics[height=85mm,width=35mm,angle=-90]{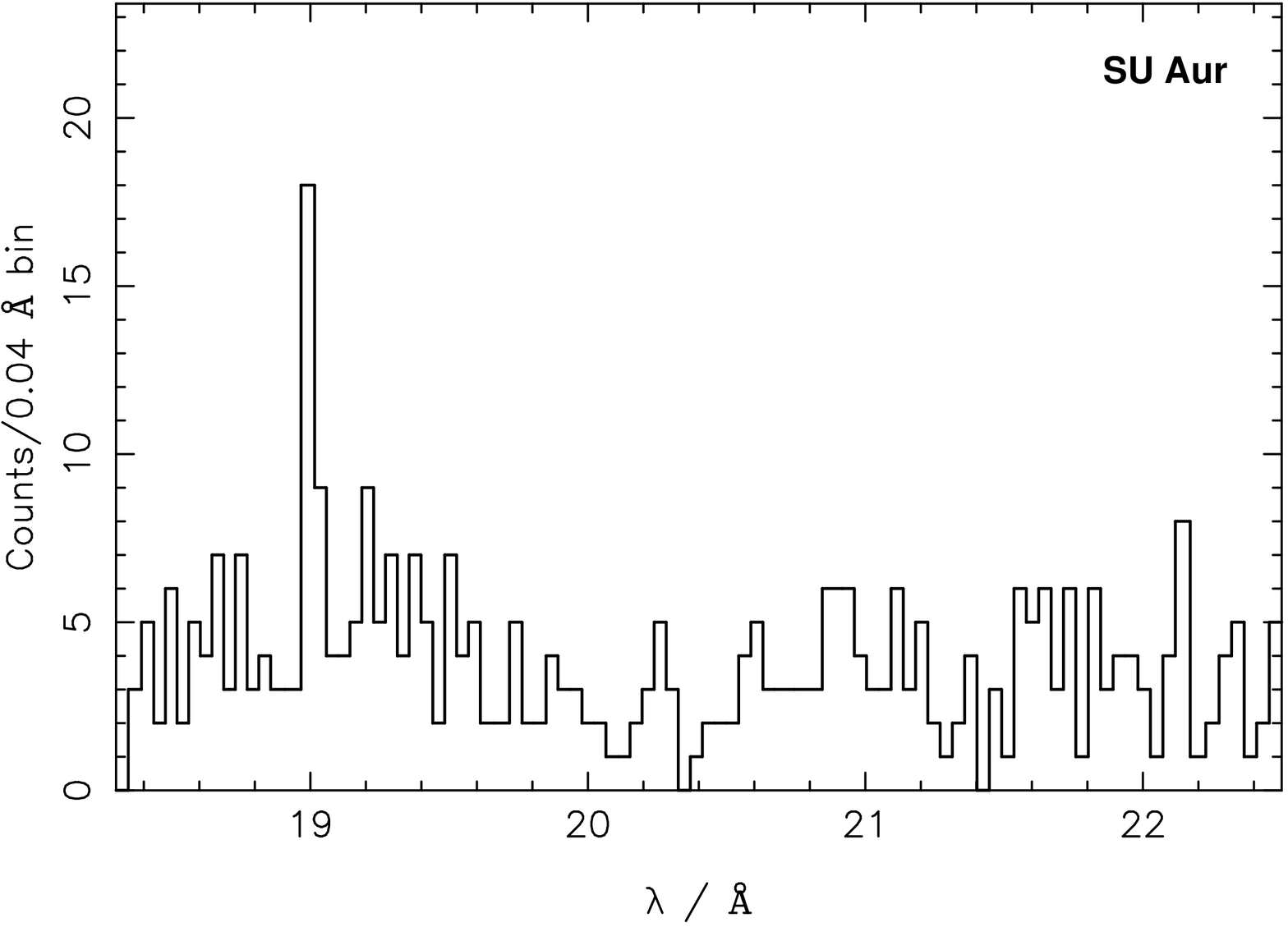}
\caption{\label{ox} Spectra of the sample CTTS covering the \ion{O}{viii} line and the \ion{O}{vii} triplet as measured 
with the RGS1 detector. Absorption increases from the top to bottom.}
\end{figure}

We used the CORA line fitting program to determine the strengths of
the resonance, intercombination and forbidden lines in the He-like triplet 
of \ion{O}{vii} (21.6, 21.8, 22.1\,\AA) and the Ly\,$\alpha$ line of \ion{O}{viii} at 18.97\,\AA \,\,as 
measured with RGS1. 
The spectra of the sample CTTS covering this wavelength-region are shown in Fig.\,\ref{ox},
the measured line counts of the \ion{O}{vii} triplet and the derived densities are given in Table\,\ref{corares}.
We used the relation $f/i =\frac{R_{0}}{1+\phi/\phi_{c}+n{e}/N_{c}}$ with f and i being the line intensities
in the forbidden and intercombination line, $R_{0}$ the low density limit of the line ratio with a adopted value of 3.95,
$N_{c}$ the critical density and $\phi/\phi_{c}$ the radiation term, which is neglected in our calculations.
Values used in the calculations were taken from \citep{pra81}; we caution that the presence of strong radiation fields
would lead us to overestimate plasma densities. 
The \ion{O}{vii} triplet traces the cool plasma around 2\,MK, is essentially free of stronger blends and therefore
well suited to investigate possible accretion scenarios, while \ion{O}{viii}
traces slightly hotter plasma around 3\,--\,5\,MK.

The analysis of the oxygen triplet of BP~Tau and its implications are presented by
\cite{bptau}, results obtained for TW~Hya are presented in \cite{twx}. 
Our analysis here is - technically -- slightly different, but we
arrive at the same results as found by \cite{twx}.
The data of CR~Cha, which unfortunately has only moderate S/N ratio and larger absorption, is analysed here for the first time.
The \ion{O}{viii} line is here clearly detected, the same applies to the
\ion{O}{vii} triplet, which, however, looks quite peculiar. Resonance and forbidden lines have about the
same strength, and the intercombination line is actually the strongest line in the triplet.  Because of the poor SNR
the errors in the line count measurements are quite substantial; also the g-ratio (f+i)/r is found
to be 2.9\,$\pm$\,2.2, but is because of its large error actually consistent with unity as theoretically expected.
The observed f/i ratio is well below unity despite its considerable measurement error, as observed for TW~Hya and
BP~Tau.

In SU~Aur the \ion{O}{vii} triplet is absent, ruling out any quantitative analysis.  We attribute the absence of
\ion{O}{vii} to the strong (circumstellar) absorption and the presence of extremely hot plasma, which increases the
continuum emission at these wavelengths.  Some weak emission may actually be present at the position of the 
oxygen triplet, but neither the intercombination nor forbidden line are clearly detected.
The \ion{O}{viii} Ly${\alpha}$ line is however clearly present 
and attenuation increases only by a factor of roughly two between
the two wavelengths. A strong \ion{O}{vii} triplet should therefore be detected in the spectra, 
even with SU~Aur's strong absorption. This indicates, that large amounts of cool plasma do not exist on SU~Aur.

\begin{table}[!ht]
\setlength\tabcolsep{5pt}
\caption{\label{corares}Measured RGS1 line counts of the \ion{O}{viii} and \ion{O}{vii}-triplet (r,i,f) lines 
and calculated densities, taking into account the effective areas at the respective wavelengths.}
{\scriptsize
\begin{tabular}{lcccc}\hline\hline
Par. & BP Tau & CR Cha & SU Aur &TW Hya\\\hline
\ion{O}{viii}& 94.5$\pm 11.4$& 36.6$\pm 7.1$ & 33.1$\pm 7.5$ & 453.0$\pm 23.8$\\
\ion{O}{vii} r & 44.9$\pm 8.4$& 7.5$\pm 4.9$ & - &234.9$\pm 17.3$\\
\ion{O}{vii} i & 34.5$\pm 7.6$  & 13.5$\pm 5.6$ & - & 113.5$\pm 12.9$\\
\ion{O}{vii} f & 11.6$\pm 5.3$ & 8.7$\pm 4.8$ & - &6.1$\pm 5.1$\\
f/i (obs.)& 0.34$\pm 0.17$ & 0.64$\pm 0.44$ & - &0.054$\pm 0.045$ \\\hline
$n_{e} (10^{11}cm^{-3})$ &3.2$^{+ 3.5}_{- 1.2}$& 1.6$^{+ 4.2}_{- 0.8}$& -&  21.1$^{+ 107}_{- 9.7}$\\ \hline
\end{tabular}
}
\end{table}

Comparing the ratio of \ion{O}{viii} to the \ion{O}{vii} line counts, we find the largest (observed) ratio for TW~Hya. 
However, absorption attenuates the \ion{O}{vii} lines more strongly
than the \ion{O}{viii} line and using our determined values for $N_{H}$, in the unabsorbed ratio the inverse trend is observed.
While typical active coronae, as well as all other sample CTTS, show a rise of the EMD in this 
temperature regime, we find that
in the accretion dominated spectrum of TW~Hya the bulk of plasma has temperatures around 2.5\,--\,3\,MK,
consistent with the results of the global fitting.

Investigating the \ion{O}{vii} triplets of our sample CTTS, 
we find that derived densities differ from those of cool main sequence stars.
While typical coronal plasma is compatible with and for very active stars no more than an order of magnitude 
above the low density limit, i.e f/i $\sim$\,2--4 and \hbox{log\,$n_{e}\sim$\,9--10} \citep{ness02},  
all CTTS with detected oxygen triplets deviate strongly from the low density limit. BP~Tau and CR~Cha 
show a density about two orders of magnitude lower, for TW~Hya it is even three orders of magnitude lower,
suggesting extremely high densities for the cool plasma of the CTTS. Alternatively, strong UV radiation fields
would explain the measured line ratios, but these fields cannot be produced by the stellar photospheres of the
underlying cool stars, therefore the radiation fields would have to be attributed to an additional hotter
component, i.e. an accretion shock.

Thus the derived densities may be considered as an upper limit, but
we point out that even in the case of UV radiation contributing to the low f/i-ratio accretion must occur,
since also a strong UV radiation points to the presence of an accretion-induced shocks.
Furthermore, the coronal plasma is likely to additionally contribute to the flux in the \ion{O}{vii}-triplet.
It was shown in \cite{bptau} that the f/i-ratio for BP~Tau is even lower, 
when only the quasi-quiescent state is considered in the analysis. 
This is plausible, because the f/i-ratio from the coronal contribution is expected to be near the low density limit and
therefore the densities in the accretion spot are underestimated in this calculation and also appear as an lower limit.
Since the size of the accretion spots or filling factors as well as the amount of coronal plasma contributing
at low temperatures are not known precisely, we just note that there are two
competing effects on the derived densities present, but they are not calling the accretion scenario in question.
While the poor data quality for CR~Cha prevents a more detailed discussion as for BP Tau, 
the derived f/i-ratio is also significantly lower than for coronal sources and the derived plasma density 
is in the same order of magnitude as it is for BP Tau. 
The derived results for CR Cha fit into the picture and support the accretion scenario for CTTS as a class. 
Therefore we conclude that the cool plasma of TW~Hya is strongly dominated by accretion, while in the cases of 
BP~Tau and CR~Cha and possibly also of SU~Aur accretion is present, but with
additional cool coronal plasma. However, the emission measure of this component is
hard to constrain given the SNR of our data. 
These findings agree well with the conclusions drawn in the previous section.

\subsubsection{Spectral changes}
\label{specch}

\begin{figure}[!ht]
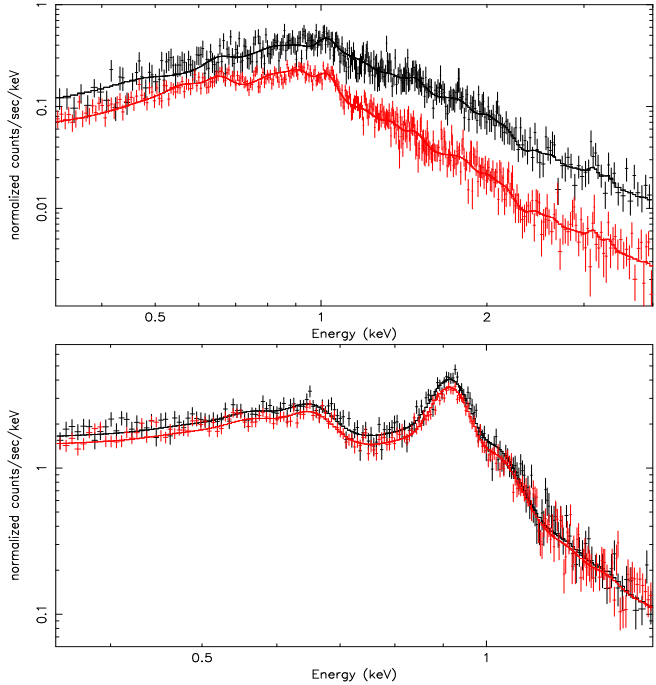

\includegraphics[width=45mm,height=86mm,angle=-90]{4247fg5a}
\includegraphics[width=45mm,height=86mm,angle=-90]{4247fg5b}
\caption{\label{hilo}Spectra of BP~Tau (top) and TW~Hya (bottom) during high (black) and low (red) state with model fits.}
\end{figure}

To investigate the spectral changes between the phases of different X-ray brightness, 
we use the more sensitive EPIC data, separated in low and high states as defined in Sect\,\ref{lcana} and
utilise the previously derived best fit models.
A comparison of the PN spectra of BP~Tau and TW~Hya during high and low state is shown in Fig.\,\ref{hilo}, 
where the high state 
represents the strong flare on BP~Tau (factor 2.4) and the period of slightly enhanced emission of TW~Hya (factor 1.1).
The differences between the spectra of the high and low state are nearly reverse for the two CTTS. 
While for BP~Tau the spectrum hardens and the
most significant changes are seen at higher energies, pointing to a coronal origin of the changes, the TW~Hya
spectra are identical at higher energies and only at lower energies the emission is enhanced, pointing to 
a higher accretion rate. The changes are more pronounced for BP~Tau due to the much higher increase in 
X-ray brightness.

\begin{table}[!ht]
\setlength\tabcolsep{6pt}  
\caption{\label{spechilo}EMD of our sample CTTS in low (left) and high (right) state as derived from PN data (MOS for SU Aur), 
$N_{H}$ in $10^{21}$cm$^{-2}$, kT in\,keV and EM in $10^{52}$cm$^{-3}$.}
\begin{center}
{\scriptsize
\begin{tabular}{lcccc}\hline\hline
Par. & \multicolumn{2}{c}{BP Tau} & \multicolumn{2}{c}{CR Cha} \\\hline
$N_{H}$ &  \multicolumn{2}{c}{1.5}  &  \multicolumn{2}{c}{2.6} \\
kT1 &  \multicolumn{2}{c}{0.18$^{+ 0.02}_{- 0.02}$}& \multicolumn{2}{c}{0.19$^{+ 0.06}_{- 0.04}$}\\
kT2 & 0.50$^{+ 0.07}_{- 0.05}$ &0.67$^{+ 0.07}_{- 0.07}$ & \multicolumn{2}{c}{0.65$^{+ 0.03}_{- 0.03}$}\\
kT3 & 2.12$^{+ 0.15}_{- 0.14}$& 2.71$^{+ 0.26}_{- 0.14}$ & 1.71$^{+ 0.20}_{- 0.19}$& 1.99$^{+ 0.23}_{- 0.14}$\\
EM1 & 4.5$^{+ 0.5}_{- 0.5}$ & 2.9$^{+ 1.0}_{- 1.0}$& 4.5$^{+ 1.5}_{- 1.2}$ & 6.3$^{+ 2.9}_{- 2.3}$\\
EM2 & 5.9$^{+ 0.4}_{- 0.5}$& 9.8$^{+ 0.9}_{- 1.2}$& 10.8$^{+ 1.5}_{- 1.7}$& 11.7$^{+ 1.4}_{- 1.9}$\\
EM3 & 6.8$^{+ 0.5}_{- 0.4}$ & 23.9$^{+ 1.4}_{- 1.7}$ & 7.7$^{+ 0.9}_{- 0.9}$& 13.3$^{+ 1.3}_{- 1.4}$\\\hline
$\chi^{2}$(dof) & \multicolumn{2}{c}{0.89 (778)}  & \multicolumn{2}{c}{1.12 (414)}\\\hline\hline
Par. & \multicolumn{2}{c}{SU Aur} & \multicolumn{2}{c}{TW Hya}\\\hline
$N_{H}$ &  \multicolumn{2}{c}{3.1}  &  \multicolumn{2}{c}{0.35} \\
kT1 &   \multicolumn{2}{c}{0.69$^{+ 0.03}_{- 0.02}$} & \multicolumn{2}{c}{0.24$^{+ 0.01}_{- 0.01}$} \\
kT2 &  \multicolumn{2}{c}{1.62$^{+ 0.09}_{- 0.07}$} & \multicolumn{2}{c}{0.60$^{+ 0.09}_{- 0.06}$}\\
kT3 &   \multicolumn{2}{c}{5.08$^{+ 1.27}_{- 0.81}$} & \multicolumn{2}{c}{1.37$^{+ 0.11}_{- 0.05}$}\\
EM1 & 10.8$^{+ 1.0}_{- 0.5}$& 11.9$^{+ 1.1}_{- 0.3}$& 21.5$^{+ 0.4}_{- 0.6}$ & 24.5$^{+ 0.4}_{- 0.8}$\\
EM2 & 21.2$^{+ 1.8}_{- 2.1}$& 29.9$^{+ 4.1}_{- 4.3}$ & 2.3$^{+ 0.8}_{- 0.4}$ & 3.2$^{+ 1.1}_{- 0.3}$ \\
EM3 & 6.9$^{+ 1.9}_{- 1.9}$& 22.7$^{+ 4.4}_{- 4.3}$ & 2.8$^{+ 0.3}_{- 0.6}$ & 2.3$^{+ 0.5}_{- 0.8}$\\\hline
$\chi^{2}$(dof) & \multicolumn{2}{c}{1.10 (929)}& \multicolumn{2}{c}{1.27 (496)}\\\hline
\end{tabular}
}
\end{center}
\end{table}

No significant changes in the elemental abundances were detected between high and low state, 
therefore we fixed strength of absorption and abundances at those values 
that were derived for the total observation and remodelled the EMD for the selected time intervals.
If we find -- using temperature as a free parameter for each state --  no significant difference in the 
temperature components, it is tied for both sets of data. 

The results of the individual modelling of the low and high states for our sample CTTS are 
summarised in Table\,\ref{spechilo}.
Very different effects on the derived temperature structures are found and also
the largest increase in emission measure is found in different temperature components for our CTTS. 
The rise in temperature and the presence of a larger amount of additional hot plasma with temperatures
around 30\,MK during the high state reflects the 
coronal origin of the increased X-ray brightness for BP~Tau. Flaring affects also the emission measure
at moderate temperatures as represented by the increased emission measure and a shift to higher temperatures,
which are in the range of 6\,--\,8\,MK.
A more moderate increase of temperatures and additional plasma at moderate temperatures but
again predominantly in the hot component
is found for CR~Cha. This also points to a coronal origin of the increased brightness 
and is consistent with the observed more moderate flare. Also for SU~Aur coronal activity is the explanation
for the increased X-ray brightness, with large amounts of additional hot plasma at temperatures of 20\,--\,50\,MK.
The flare plasma is extremely hot,
but since large amounts of extremely hot plasma are already present during quasi-quiescence, 
no significant increase in temperature accompanies the brightening.
The coolest detected component of SU~Aur, which already has temperatures around 8\,MK, 
is again not strongly affected by the flaring. 
In contrast, none of these signatures are connected with the flux increase for TW~Hya. 
No differences in the temperature structure were found and the additional plasma
is found mainly in the coolest temperature component, supporting the scenario of increased accretion.
With 2.5\,--\,3\,MK this component is much cooler than typical flare plasma temperatures, but 
fits to the temperatures as expected for plasma generated in an accretion spot.

\section{Summary and Conclusions}
\label{summ}
 
We have presented the first comparative study using high and medium resolution X-ray spectra of
classical T~Tauri stars observed with the new generation X-ray telescopes so far. The results derived from these data are 
complementary to the ones obtained from the large sample of COUP sources, since
high-resolution spectroscopy and low-energy sensitivity additionally permits the investigation of the cool plasma components.

Using the XMM-Newton observations of BP~Tau, CR~Cha, SU~Aur and TW~Hya we determined X-ray properties of 
accreting young pre-main sequence stars.
Two different mechanisms are likely to contribute to the production of 
X-ray emission in those objects, first coronal, i.e. magnetic activity, and second, magnetically funneled accretion.
We investigate variability, global spectra and density sensitive lines 
to check for the presence and relative contribution of these two mechanisms for the individual objects.
For this purpose we utilize the emission measure distribution and its changes, specific abundance pattern 
and density diagnostics. Interpreting very high densities as indicator for accretion shocks 
and a high temperature component as indicator for coronal activity, we derive the following conclusions
which are likewise supported by the analysis of abundance analysis and spectral variability.
In all targets where the \ion{O}{vii}-triplet is observable density analysis leads to higher densities than 
observed in any pure coronal source. Additionally, a high temperature component is present in all targets.
We therefore argue that both X-ray generating processes are present in our sample CTTS, but at very different levels
and importance in individual objects.

The X-ray emission of BP~Tau is overall dominated by coronal activity and a stronger flare of again
probably coronal origin is observed.
Even in quasi-quiescent phases the emission measure distribution is dominated by medium and hot temperature plasma 
whereof significant amounts are present at temperatures around 20\,MK. 
However, the analysis of density sensitive lines
confirms that accretion is also present and actually dominates the cool plasma with temperatures of a few MK. 
CR~Cha is similar to BP~Tau but more strongly absorbed and the coronal plasma temperatures are slightly lower.
It also appears to be dominated by the coronal contribution,
likewise there is evidence for high density plasma at cool temperatures.
SU~Aur is the by far brightest and with temperatures of at least up to 50\,MK hottest X-ray source in our sample.
A powerful and active corona is the essential contributor to its X-ray spectrum and frequent flaring is observed. 
Unfortunately the very strong absorption prevents definite conclusions about the cooler plasma component.
TW~Hya is the other hand strongly dominated by accretion, but an additional coronal component is clearly detected.
It is the prototype and still the outstanding and only example of an accretion dominated star.
Most of its plasma is at cool temperatures typical for 
accretion spots and it exhibits by far the lowest \ion{O}{vii} f/i ratio. Also the highest neon to oxygen ratio,
which may be interpreted via depletion of grain forming element, is found for TW~Hya. This is
plausible since TW~Hya is the oldest and most evolved object in our sample. 
In the other sample CTTS neon is also enhanced, but
the abundance pattern is more reminiscent of the inverse FIP effect found in active stars.

Considering the global picture of stellar evolution towards the main sequence,
we find that magnetic processes play a major role in high energy phenomena in all our sample CTTS. 
While the specific stellar evolution and its timescale might depend on e.g. spectral type,
the generation of X-rays via accretion and coronal activity is apparently a common feature of CTTS in general.

\begin{acknowledgements}
This work is based on observations obtained with XMM-Newton, an ESA science
mission with instruments and contributions directly funded by ESA Member
States and the USA (NASA).\\
This research has made use of the SIMBAD database, operated at CDS, Strasbourg, France.
(http://simbad.u-strasbg.fr)\\
J.R. acknowledges support from DLR under 50OR0105.

\end{acknowledgements}


\begin{thebibliography}{}
\bibitem[Anders \& Grevesse(1989)]{and89}Anders, E., Grevesse, N. 1989, Geo- et Cosmochimica Acta, 53, 197
\bibitem[Argiroffi et al.(2005)]{twa5}Argiroffi, C., Maggio, A., Peres, G., et al. 2005, A\&A, 439, 1149
\bibitem[Arnaud(1996)]{xspec}Arnaud, K.A. 1996, ASP Conf. Series, 101, 17
\bibitem[Asplund(2005)]{asp05}Asplund, M. 2005,  ASP Conf. Series, 336, 25
\bibitem[Calvet \& Gullbring(1998)]{cal98}Calvet, N., Gullbring, E. 1998, ApJ, 509, 802
\bibitem[DeWarf et al.(2003)]{dew03}DeWarf, L.E., Sepinsky, J.F., Guinan, E.F., et al. 2003, ApJ, 590, 357
\bibitem[Drake et al.(2005)]{dra05}Drake, J.J., Testa, P., Hartmann, L. 2005, ApJ, 627, L149
\bibitem[Ehle et al.(2003)]{xmm}Ehle, M., Breitfellner, M., Gonzales Riestra, M., et al. 2003, XMM-Newton User's Handbook
\bibitem[Favata et al.(2005)]{fav05}Favata, F., Flaccomio, E., Reale, F., et al. 2005, ApJS, 160, 469
\bibitem[Feigelson \& DeCampli(1981)]{fei81}Feigelson, E.D., DeCampli, W.M. 1981, ApJ, 243, L89
\bibitem[Feigelson \& Kriss(1989)]{fei89}Feigelson, E.D., Kriss, G.A. 1989, ApJ, 338, 262
\bibitem[Feigelson et al.(1993)]{fei93}Feigelson, E.D., Casanova, S., Montmerle, T., Guibert, J. 1993, ApJ, 416, 623
\bibitem[Feigelson \& Montmerle(1999)]{fei99}Feigelson, E.D., Montmerle, T. 1999, Ann. Rev. A\&A, 37, 363
\bibitem[Feigelson et al.(2003)]{fei03}Feigelson, E.D., Gaffney, J.A., Garmire, G., et al. 2003, ApJ, 584, 911
\bibitem[Grevesse \& Sauval(1998)]{grev98}Grevesse, N., Sauval, A.J. 1998, Space Sci. Rev., 85, 161
\bibitem[Gauvin \& Strom(1992)]{gau92}Gauvin, L.S. and Strom, K.M. 1992, ApJ, 385,217
\bibitem[G\"udel et al.(2005)]{gue05}G\"uedel, M., Skinner, S.L., Briggs, K.R., et al. 2005, ApJ 626, L53
\bibitem[Gullbring et al.(1997)]{gul97}Gullbring, E., Barwig, H., Schmitt, J.H.M.M. 1997, A\&A, 324, 155
\bibitem[Gullbring et al.(1998)]{gul98}Gullbring, E., Hartmann, L., Briceno, C., Calvet, N. 1998, ApJ, 492, 323
\bibitem[Hartigan et al.(1995)]{har95}Hartigan, P., Edwards, S., Ghandour, L. 1995, ApJ, 452, 736
\bibitem[Herczeg et al.(2004)]{twuv}Herczeg, G.J., Wood, B.E., Linsky, J.L., et al. 2004, ApJ, 607, 369
\bibitem[Kastner et al.(1999)]{kas99}Kastner, J.H., Huenemoerder, D.P., Schulz, N.S., et al. 1999, ApJ, 525, 837
\bibitem[Kastner et al.(2002)]{twc}Kastner, J.H., Huenemoerder, D.P., Schulz, N.S., et al. 2002, ApJ, 567, 434
\bibitem[Kirsch et al.(2004)]{kir04}Kirsch, M.G.F., Altieri, B., Chen, B., et al. 2004, astro-ph/0407257
\bibitem[Loiseau et al.(2004)]{sas}Loiseau, N. (edt.), Ehle, M., Pollock, A.M.T., et al. 2004, User's Guide to XMM-Newton Science Analysis System Issue 3.1
\bibitem[Makarov \& Fabricius(2001)]{mak01}Makarov, V.V., Fabricius, C. 2001, A\&A, 368, 866
\bibitem[Meeus et al. (2003)]{meeus03} Meeus, G., Sterzik, M., Bouwman, J., Natta, A. 2003, A\&A, 409, L25
\bibitem[Muzerolle et al.(2003)]{muz03}Muzerolle, J., Calvet, N., Hartmann, L, D'Alessio, P. 2003, ApJ, 597, L149
\bibitem[Natta et al.(2000)]{nat00}Natta, A., Meyer, M.R., Beckwith, S.V.W. 2000, ApJ, 534, 838
\bibitem[Ness et al.(2002)]{ness02}Ness, J.-U., Schmitt, J.H.M.M., Burwitz, V., et al. 2002, A\&A, 394, 911
\bibitem[Ness \& Wichmann(2002)]{cora}Ness, J.-U., Wichmann, R. 2002, AN, 323, 129
\bibitem[Neuh\"auser et al.(1995)]{neu95}Neuh\"auser, R., Sterzik, M.F., Schmitt, J.H.M.M., et al. 1995, A\&A, 297, 391
\bibitem[Pallavicini et al.(2004)]{pal04}Pallavicini, R., Franciosini, E., Randich, S. 2004, MSAI, 75, 434 
\bibitem[Pradhan \& Shull(1981)]{pra81}Pradhan, A.K., Shull, J.M. 1981, ApJ, 249, 821
\bibitem[Preibisch et al.(2005)]{pre05}Preibisch, T., Kim, Y.-C., Favata, F., et al. 2005, ApJS, 160, 401
\bibitem[Reipurth et al.(1996)]{rei96}Reipurth, B., Pedrosa, A., Lago, M.T.V.T. 1996, A\&AS, 120, 229
\bibitem[Robrade \& Schmitt(2005)]{mdwarfs}Robrade, J., Schmitt, J.H.M.M. 2005, A\&A, 435, 1073
\bibitem[Schmitt et al.(2005)]{bptau}Schmitt, J.H.M.M., Robrade, J., Ness, J.-U., et al. 2005, A\&A, 432, L35
\bibitem[Shu et al.(1994)]{shu94}Shu, F., Najita, J., Ostriker, E., et al. 1994, ApJ, 429, 781
\bibitem[Smith et al.(2001)]{apec}Smith, R.K., Brickhouse, N.S., Liedahl, D.A., Raymond, J.C. 2001, ASP Conf. Series, 247, 161
\bibitem[Stelzer \& Neuh\"auser(2001)]{ste01}Stelzer, B., Neuh\"auser, R. 2001, A\&A, 377, 538
\bibitem[Stelzer \& Schmitt(2004)]{twx}Stelzer, B., Schmitt, J.H.M.M. 2004, A\&A, 418, 687
\bibitem[Walter \& Kuhi(1981)]{wal81}Walter, F.M., Kuhi, L.V. 1981, ApJ, 250, 254
\bibitem[Wichmann et al.(1998)]{wich98}Wichmann, R., Bastian, U., Krautter, J., et al. 1998, MNRAS, 301, L39
\bibitem[Wilner et al.(2005)]{wil05}Wilner, D.J., Alessio, P.D., Calvet, N., et al. 2005, ApJ, 626, L109
\bibitem[Zuckerman et al.(2001)]{zuck01}Zuckerman, B., Webb, R.A., Schwartz, M., Becklin, E.E. 2001, ApJ, 549, L233
\end{thebibliography}
\end{document}